\documentclass{appolb}
\usepackage{epsfig}
\newcommand{\GeV}{\,\mathrm{GeV}}

\newcommand{\ipb}{\,\mathrm{pb}^{-1}}
\newcommand{\etoeg}{e^* \to e\gamma}
\newcommand{\etoeZtoeqq}{e^* \to eZ \to e q\bar{q}}
\newcommand{\etoeqq}{e^* \to e q\bar{q}}
\newcommand{\etovWtovqq}{e^* \to \nu W \to \nu q\bar{q}'}
\newcommand{\etovqq}{e^* \to \nu q\bar{q}'}
\newcommand{\vtovg}{\nu^* \to \nu\gamma}
\newcommand{\vtovZtovqq}{\nu^* \to \nu Z \to \nu q\bar{q}}
\newcommand{\vtovqq}{\nu^* \to \nu q\bar{q}}
\newcommand{\vtoeWtoeqq}{\nu^* \to eW \to e q\bar{q}'}
\newcommand{\vtoeqq}{\nu^* \to e q\bar{q}'}

\newcommand{\ptmiss}{\not\kern-0.5ex{P_T}}

\newcommand{\sigbr}{\sigma\times{\rm BR}}
\newcommand{\fovl}{f/\Lambda}
\begin{document}
\title{{\bf\boldmath Search for excited fermions in $ep$ collisions at HERA}
\thanks{Presented at the X International Workshop on Deep Inelastic Scattering,
  Cracow, Poland, April 30 -- May 4, 2002}
} \author{Ainas Weber \address{Physikalisches Institut der Universit\"at Bonn\\
  Nussallee 12, D-53115 Bonn, Germany } } \maketitle
\begin{abstract}
  Heavy excited electrons and neutrinos have been sought by the H1 and ZEUS
  experiments at HERA. For the $e^*$ ($\nu^*$) searches, $120\ipb$ ($16\ipb$) of
  $ep$ collision data have been analysed. No evidence for any excited lepton has
  been found, and limits on the characteristic couplings have been derived.
\end{abstract}
%
\section{Introduction}

The observation of heavy excited states of fermions would indicate that these
particles are composite rather than elementary. In high-energy $ep$ collisions,
such excited states could be produced directly, with masses up to the
centre-of-mass energy of the collider. 

The recent excited-fermion searches performed by the H1 and ZEUS experiments
have focussed on $e^*$ and $\nu^*$~\cite{h1nustar,h1estar,zeusfstar,zeusestar}.
Excited states of electrons were sought using the following data sets, where the
centre-of-mass energies and the approximate integrated luminosities per
experiment are given in brackets: 1994--1997 $e^+p$ ($300\GeV$, $40\ipb$),
1998--1999 $e^-p$ ($318\GeV$, $16\ipb$), 1999--2000 ($318\GeV$, $66\ipb$). The
decay channels considered are $\etoeg$, $\etovWtovqq$ and $\etoeZtoeqq$.

Excited states of neutrinos were sought in the $e^-p$ data sets only, since
the $\nu^*$ production cross section for masses beyond $200\GeV$ in $e^-p$
collisions is two orders of magnitude higher than that in $e^+p$. The decay
channels analysed are $\vtovg$, $\vtoeWtoeqq$ and $\vtovZtovqq$.

The simulation of $e^*$ and $\nu^*$ signal events was based on the
phenomenological compositeness model by Hagiwara et al.~\cite{hagiwara}. For estimating
the backgrounds from Standard Model processes, Monte Carlo samples of NC and CC
DIS as well as of photoproduction (PHP) and QED-Compton events were employed.

\section{Searches for excited electrons}

\subsection{$\etoeg$}

Among the $e^*$ channels considered, the photonic decay mode, $\etoeg$,
provides the most striking experimental signature, featuring two isolated
electromagnetic clusters with large transverse energies. Typically, no further
activity is observed in the detector, except for possible energy deposits by the
proton remnant around the forward beampipe. As for the backgrounds,
NC DIS and QED-Compton events contribute with similar magnitudes. 

\subsection{$\etovWtovqq$}

This final state is characterised by a large amount of missing transverse
momentum, $\ptmiss$, due to the neutrino escaping undetected as well as
by two hadronic jets from the $W$ decay. The relevant background processes are
predominantly multi-jet CC DIS and, less pronounced, PHP. The H1 search required
the identification of two jets, whereas ZEUS employed cuts on global event
variables like transverse hadronic energy and hadronic mass. Both experiments
applied an electron veto.

\subsection{$\etoeZtoeqq$}

Due to the hadronic $Z$ decay, this final state features two jets with high
transverse energy. In addition, there is a forward-going high-energy
electron. The only relevant source of background is constituted by NC
DIS. Again, the two experiments used different approaches in treating the
hadronic final state, namely on the basis of jets (H1) and global event
variables (ZEUS), respectively.

\section{Searches for excited neutrinos}

\subsection{$\vtovg$}

The photonic $\nu^*$ decay gives rise to a particularly rare experimental
signature: one isolated, high-energy electromagnetic cluster in the forward
direction plus a large amount of $\ptmiss$. Background arises
predominantly from CC DIS.

\subsection{$\vtoeWtoeqq$ and $\vtovZtovqq$}

The topologies of the $e q\bar{q}'$ and $\nu q\bar{q}'$ final states originating
from $\nu^*$ decays are similar to the ones of the corresponding $e^*$ final
states. Thus, similar selection criteria as in the corresponding $e^*$ channels
were employed by the two experiments.

\section{Results}

The numbers of candidate and background events obtained by the sear\-ches are
summarised in Table~\ref{tab-events}. No excess of data events over the expected
background has been observed in either of the decay channels analysed. Thus,
upper limits at $95\,\%$ confidence level have been set on the cross section
times the branching ratio, $\sigbr$, and on the coupling over the compositeness
scale, $\fovl$. The latter limits require as input branching ratios and cross
sections from the specific model used.

\begin{figure}
\raisebox{3mm}{
\epsfig{file=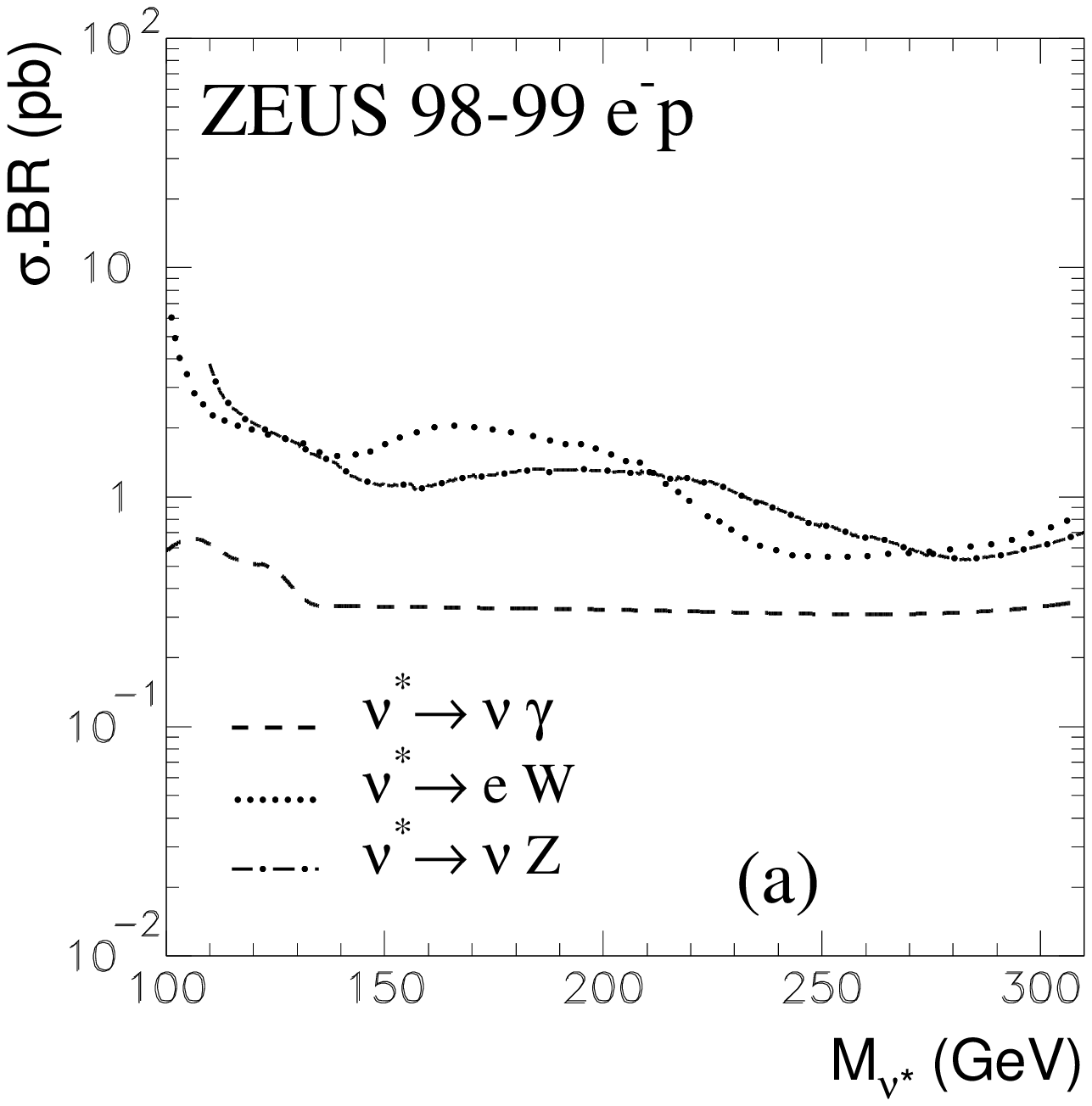,width=0.47\linewidth}}\hfill
\epsfig{file=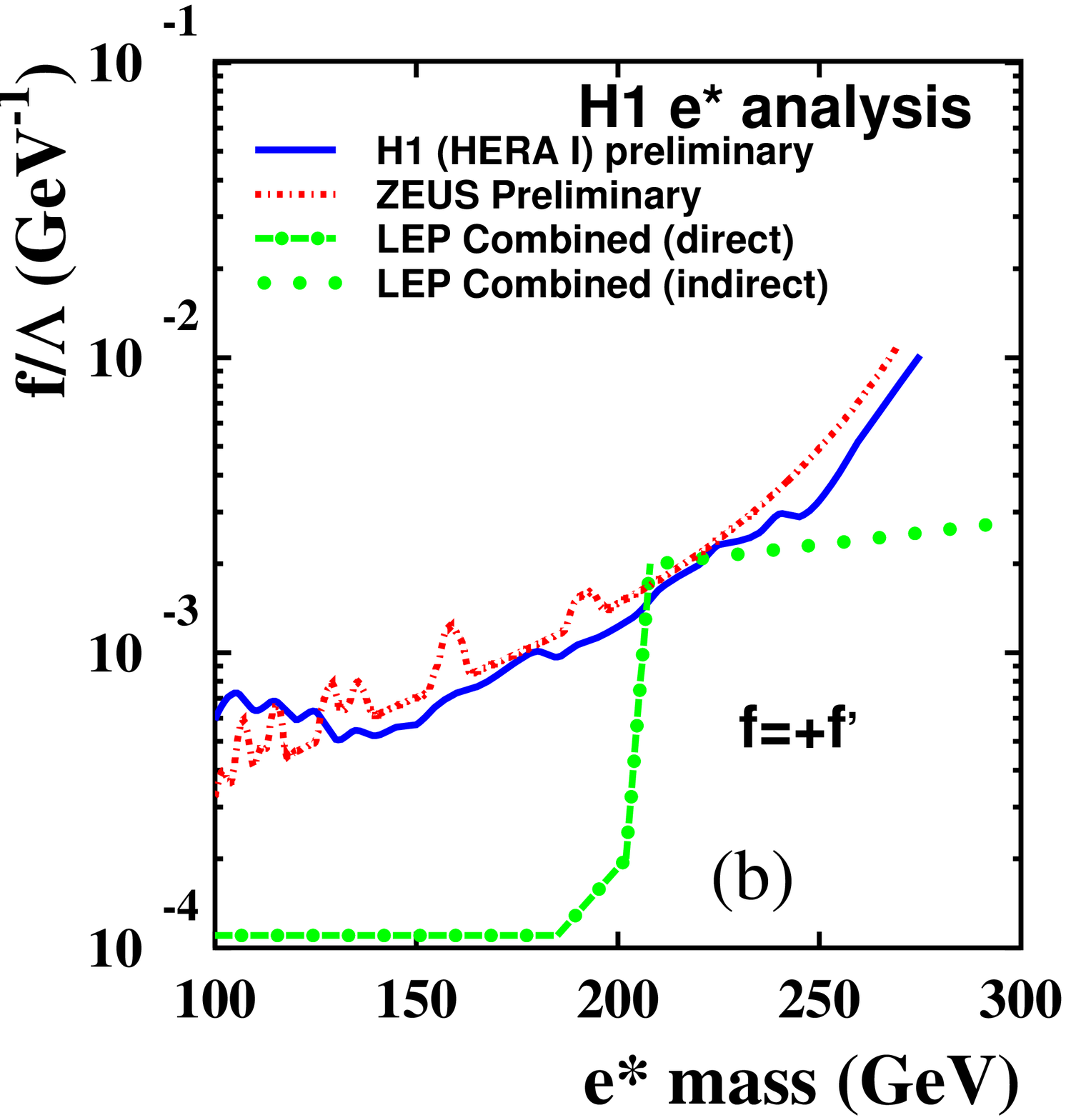,width=0.49\linewidth}
\caption{(a) Upper limits at $95\,\%$ confidence level on $\sigbr$ as a function
  of the $\nu^*$ mass~\protect\cite{zeusfstar}. (b) Upper limits on $\fovl$ for
  $e^*$. The H1 (ZEUS) limits are based on $120\ipb$ ($82\ipb$) of $ep$ data.
  For comparison, the corresponding limits by the LEP experiments are
  shown~\protect\cite{lep}.}
\label{fig-lim1}
\end{figure}

\begin{figure}
\epsfig{file=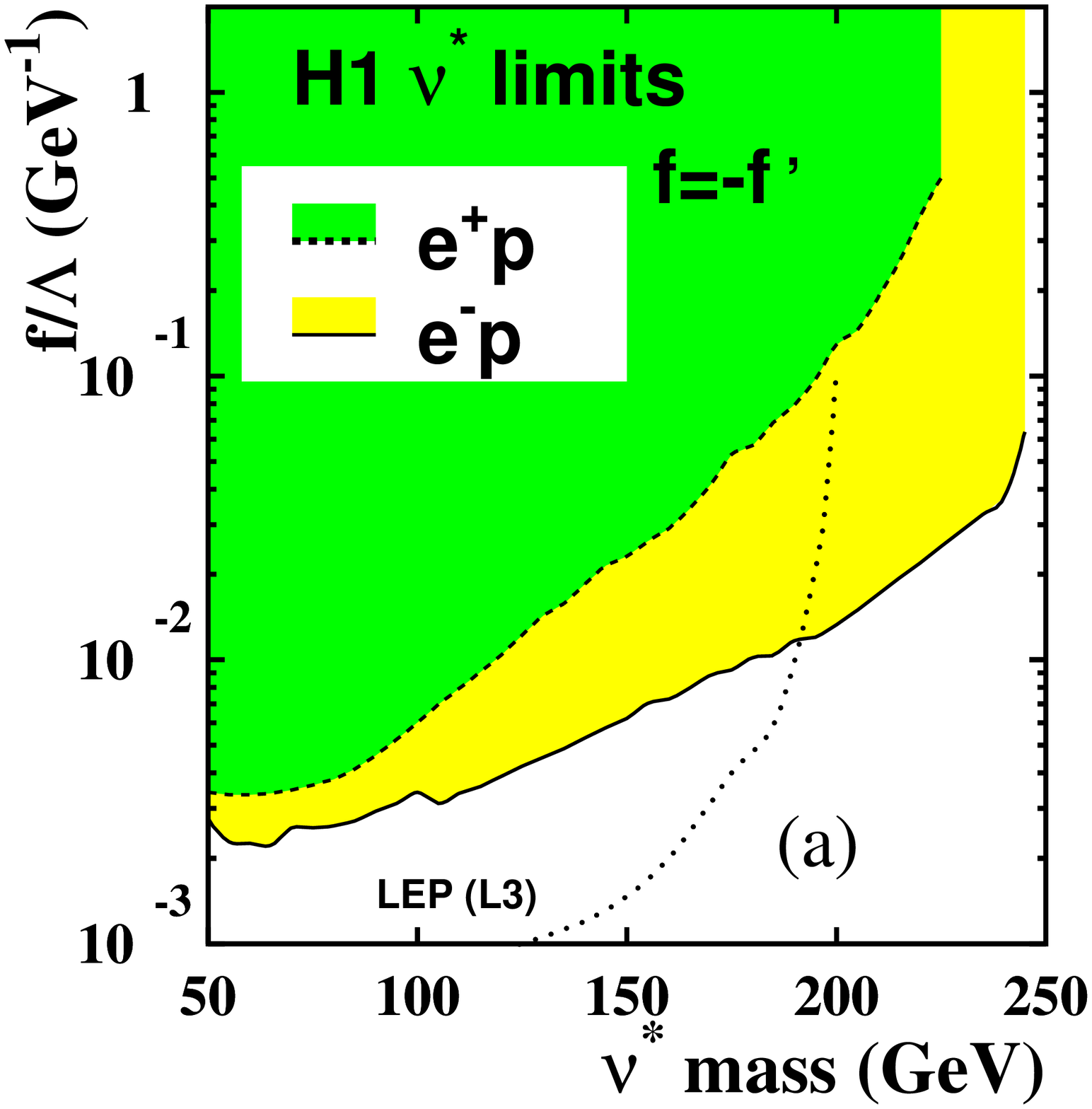,width=0.49\linewidth}\hfill
\epsfig{file=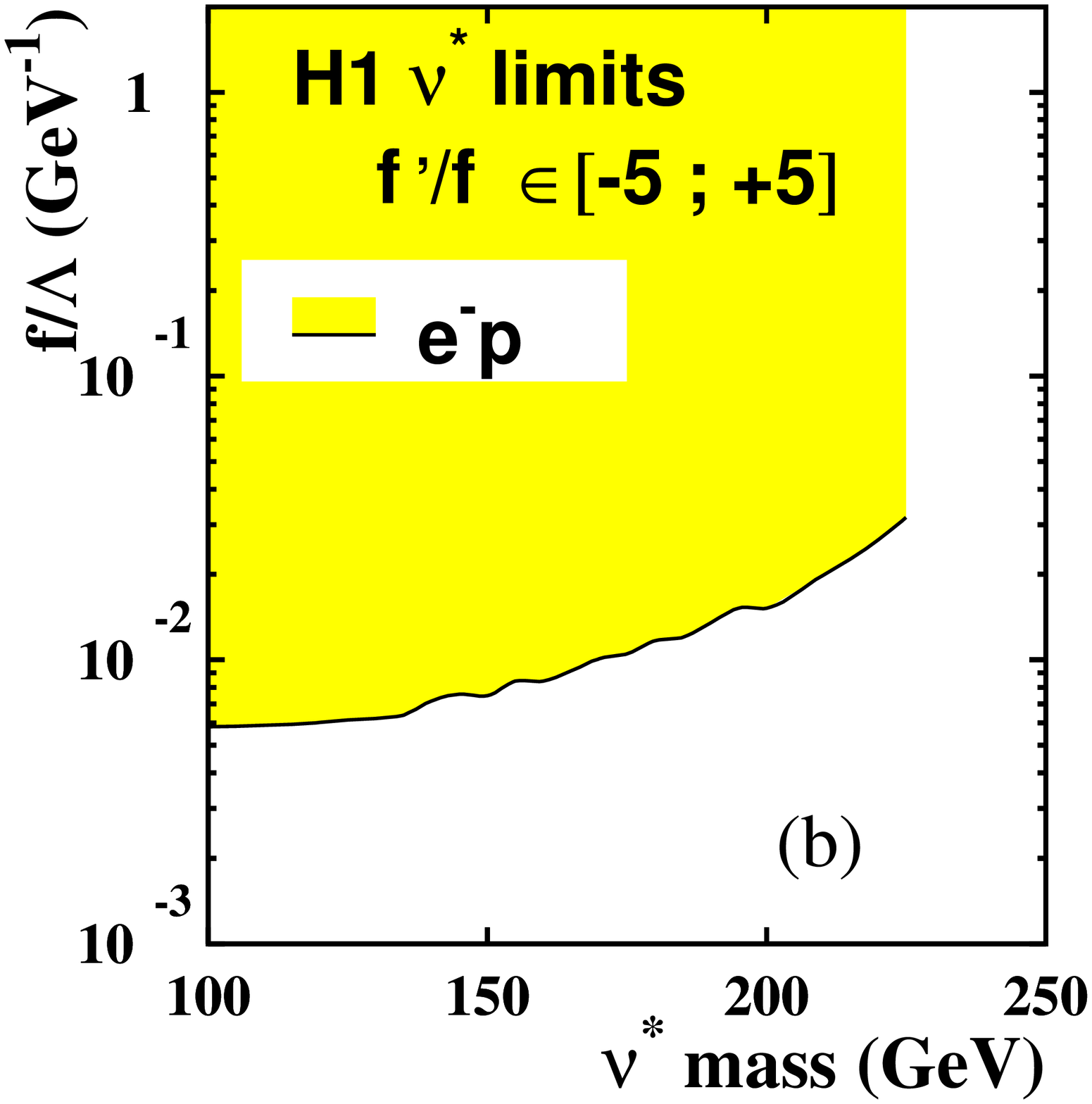,width=0.49\linewidth}
\caption{(a) Upper limits on $\fovl$ for $\nu^*$ assuming $f=-f'$. The
  $e^+p$~\protect\cite{h1fstar} and $e^-p$~\protect\cite{h1nustar} limits are
  based on integrated luminosities of $37\ipb$ and $16\ipb$, respectively. The
  $e^-p$ limits are more stringent, though, due to the higher $\nu^*$ production
  cross-section compared to $e^+p$ collisions. (b) Upper limits on $\fovl$ for $\nu^*$,
  not depending on the ratio $f/f'$~\protect\cite{h1nustar}.}
\label{fig-lim2}
\end{figure}

In Fig.~\ref{fig-lim1}~(a), the ZEUS upper limits on $\sigbr$ for excited
neutrinos are displayed. In Fig.~\ref{fig-lim1}~(b), $\fovl$ limits for excited
electrons are shown. In all limit plots, the areas above the curves are
excluded. The H1 curve in Fig.~\ref{fig-lim1}~(b) has been derived from
combining the three decay channels considered, using the conventional assumption
$f=f'$ for the coupling constants in the Hagiwara model; the ZEUS curve is from
$\etoeg$ only. In the $\nu^*$ case, $\fovl$ limits are usually derived for both
assumptions $f=f'$ and $f=-f'$. The H1 limits for the latter assumption,
allowing for photonic $\nu^*$ decays, are shown in Fig.~\ref{fig-lim2}~(a).

In addition, H1 has derived less model-dependent $\fovl$ limits for $\nu^*$. At
each mass point, the ratio of the coupling constants was varied in the range
$-5<f'/f<5$, the corresponding $\fovl$ limit was calculated, and the worst of
these limit values was chosen. The resulting curve is displayed in
Fig.~\ref{fig-lim2}~(b). 

In comparison with existing limits by H1 and ZEUS, the new $\sigbr$ and $\fovl$
limits for $e^*$ and $\nu^*$ are more stringent and extend towards higher
masses. For $\nu^*$, with no indirect limits available from the LEP experiments,
H1 and ZEUS set the most stringent limits in the high-mass region.

\begin{table}
\begin{center}
\begin{tabular}{|c|c||c|c||c|c|}\hline
decay    & data   & \multicolumn{2}{c||}{H1} & \multicolumn{2}{c|}{ZEUS} \\ \cline{3-6} 
channel  & sample & events & background & events & background\\
\hline\hline
           & 94--97 & 8   & 7.2$\pm$1.0$\pm$0.1  &  18   & 20.1$\pm$1.2 \\
$\etoeg$   & 98--99 & 4   & 4.0$\pm$0.7$\pm$0.2  &  10   & 8.7 \\
           & 99--00 & 12  & 15.6$\pm$1.7$\pm$0.4 &  22   & 30.8  \\
\hline
           & 94--97 & 2   & 2.4$\pm$0.2$\pm$0.7  &  13  & 13.9$\pm$1.1 \\
$\etovqq$  & 98--99 & 5   & 3.9$\pm$0.2$\pm$0.7  &      &       \\
           & 99--00 & 8   & 6.1$\pm$0.4$\pm$1.5  &      &         \\
\hline
           & 94--97 & 6   & 7.1$\pm$2.1$\pm$2.8  & 32   & 32.9$\pm$1.1  \\
$\etoeqq$  & 98--99 & 4   & 5.6$\pm$0.4$\pm$1.2  &      &      \\
           & 99--00 & 31  & 25.3$\pm$1.9$\pm$5.5 &      &        \\
\hline\hline
$\vtovg$   & 98--99 & 2   & 3.0$\pm$0.2$\pm$1.2  & 2    & 1.5$\pm$0.2 \\
\hline
$\vtoeqq$  & 98--99 & 6   & 7.0$\pm$0.6$\pm$1.4  & 20   & 15.0$\pm$1.3 \\
\hline
$\vtovqq$  & 98--99 & 1   & 3.7$\pm$0.2$\pm$0.9  & 16   & 13.5$\pm$0.6 \\
\hline
\end{tabular}
\caption{Numbers of observed events and expected backgrounds in the six decay
  channels, given separately for the different data samples analysed by the two
  experiments. The errors stated are statistical and systematic for H1, and
  statistical for ZEUS.}
\label{tab-events}
\end{center}
\end{table}

\begin{thebibliography}{99}

\bibitem{h1nustar}  H1 Coll., C.~Adloff~et~al., Phys. Lett. B~525 (2002) 9. 

\bibitem{h1estar} H1 Coll., H1-prelim-02-061, Feb.~2002.

\bibitem{zeusfstar} ZEUS Coll., S.~Chekanov~et~al., Preprint DESY~01-132, 2001.

\bibitem{zeusestar} ZEUS Coll., Abstract 607, EPS01, Budapest, 2001. 

\bibitem{hagiwara} K.~Hagiwara, S.~Komamiya and D.~Zeppenfeld, Z. Phys. C~29 (1985) 115;\\
U.~Baur, M.~Spira and P.M.~Zerwas, Phys. Rev. D~42 (1990) 815;\\
F.~Boudjema, A.~Djouadi and J.L.~Kneur, Z. Phys. C~57 (1993) 425.

\bibitem{h1fstar}  H1 Coll., C.~Adloff~et~al., Eur. Phys. J. C~17 (2000) 567.

\bibitem{lep} LEP Exotica Working Group, Contributions to EPS01 and LP01,
  http://lepexotica.web.cern.ch/LEPEXOTICA.

\end{thebibliography}
\end{document}